# Observation of a multitude of correlated states at the surface of bulk 1T-TaSe$_2$ crystals

Yi Chen[1,2,3,4,†], Wei Ruan[1,2,5,†], Jeffrey D. Cain[1,2,6], Ryan L. Lee[1,7], Salman Kahn[1,2], Caihong Jia[1,8], Alex Zettl[1,2,6], Michael F. Crommie[1,2,6,*]

[1]Department of Physics, University of California, Berkeley, California 94720, USA

[2]Materials Sciences Division, Lawrence Berkeley National Laboratory, Berkeley, California 94720, USA

[3]International Center for Quantum Materials, School of Physics, Peking University, Beijing 100871, China 6

[4]Collaborative Innovation Center of Quantum Matter, Beijing 100871, China

[5]State Key Laboratory of Surface Physics and Department of Physics, Fudan University, Shanghai 200438, China

[6]Kavli Energy Nano Sciences Institute at the University of California Berkeley and the Lawrence Berkeley National Laboratory, Berkeley, California 94720, USA

[7]Joseph Henry Laboratories and Department of Physics, Princeton University, Princeton, New Jersey 08544, USA

[8]Henan Key Laboratory of Photovoltaic Materials and Laboratory of Low-dimensional Materials Science, School of Physics and Electronics, Henan University, Kaifeng 475004, China

† These authors contributed equally to this work.

*e-mail: crommie@berkeley.edu

**Abstract**

The interplay between electron-electron interactions and structural ordering can yield exceptionally rich correlated electronic phases. We have used scanning tunneling microscopy to investigate bulk 1T-TaSe$_2$ and have uncovered surprisingly diverse correlated surface states thereof. These surface states exhibit the same in-plane charge density wave ordering but dramatically different electronic ground states ranging from insulating to metallic. The insulating variety of surface state shows signatures of a decoupled surface Mott layer. The metallic surface states, on the other hand, exhibit zero-bias peaks of varying strength that suggest Kondo phases arising from coupling between the Mott surface layer and the metallic bulk of 1T-TaSe$_2$. The surface of bulk 1T-TaSe$_2$ thus constitutes a rare realization of the periodic Anderson model covering a wide parameter regime, thereby providing a model system for accessing different correlated phenomena in the same crystal. Our results highlight the central role played by strong correlations in this material family.

# I. Introduction

In layered transition-metal dichalcogenides such as 1T-TaSe$_2$ and 1T-TaS$_2$, the formation of a star-of-David charge density wave (CDW) creates a half-filled flat band that would cross the Fermi level ($E_F$) in the absence of electron correlations and interlayer coupling [1-5]. In the actual materials, however, both electron correlations and interlayer coupling can play significant roles, and their separate influences are often entangled [6-8]. In 1T-TaS$_2$, for example, the insulating bulk phase has long been believed to originate from electron correlations [9,10], but recent evidence suggests that interlayer coupling effects can cause it to be a band insulator [7,8]. In TaS$_2$'s less-studied sister material, 1T-TaSe$_2$, isolated single layers were found to be Mott insulators [5] that exhibit quantum spin liquid behavior [11,12]. Interlayer coupling, however, was shown to rapidly quench the insulating state for bilayer and trilayer 1T-TaSe$_2$ [5], consistent with the transition of 1T-TaSe$_2$ to a bulk metal with no obvious strong correlation effects [13,14]. The role of electron interaction effects thus remain unclear for the bulk forms of this material family and debate continues as to whether they can host exotic correlated phenomena [15-17].

The surface of bulk 1T-TaSe$_2$ provides a window into this behavior, but has so far yielded somewhat confusing results. Scanning tunneling microscopy and spectroscopy (STM/STS) studies on different 1T-TaSe$_2$ samples have shown inconsistent surface states, ranging from metallic [18-20] to insulating [18,21,22]. This diverse behavior was attributed to sample quality variations [22] but recently suggested to be related to intrinsic interlayer CDW stacking orders [23]. On the other hand, electron correlation effects also appear to be relevant at the surface of 1T-TaSe$_2$. The insulating surface state was confirmed early on by angle-resolved photoemission spectroscopy and was interpreted as a surface Mott insulating state arising from CDW-induced band narrowing that causes the electron-electron repulsion, $U,$ to exceed the bandwidth, $W$, at the surface [21,22]. How to reconcile the very different surface states identified in 1T-TaSe$_2$ and achieve a coherent understanding of their nature remains an elusive goal despite 30 years of studies.

Here we present an experimental study that reveals electron correlations to be the universal origin of the different 1T-TaSe$_2$ surface states. Our STM measurements show that the same high-quality bulk 1T-TaSe$_2$ crystal can host multiple surface states that all exhibit the same in-plane star-of-David CDW pattern, but show different correlated ground states ranging from metallic to insulating within each macroscopic cleavage domain. The metallic surface states exhibit zero-bias peaks in STS (i.e., at the Fermi energy, $E_F$) having varying strength. STM spectroscopic mapping reveals that the low-energy electronic structure for the different surface states is always dominated by the same CDW orbital located at the center of the CDW unit cell. The correlated nature of these surface states is seen foremost in the insulating case which shares a striking similarity to the Mott insulator ground state of single-layer 1T-TaSe$_2$. The narrow zero-bias peaks of the metallic 1T-TaSe$_2$ surface states, on the other hand, suggest that Kondo resonances arise due to coupling of surface magnetic moments to underlying metallic bulk layers. The existence of surface magnetic moments implied by the presence of the Kondo resonance provides strong evidence for the underlying "Mottness" of 1T-TaSe$_2$.

## II. Experimental Setup

Bulk 1T-TaSe$_2$ was grown by chemical vapor transport at 950°C using iodine as a transport agent [24]. After growth, the 1T-TaSe$_2$ crystals were rapidly quenched to room temperature in ice water to maintain the 1T structure. The bulk 1T-TaSe$_2$ was cleaved in UHV at room temperature before being transferred to the STM without breaking vacuum. A typical optical image of cleaved bulk 1T-TaSe$_2$ is shown in the inset to Fig. 1(b). Large, flat patches of the surface are visible.

**III. Results**

All of the bulk 1T-TaSe$_2$ surfaces that we scanned showed similar STM topography at typical scanning parameters as shown in Figs. 1(b)-1(d). Here the triangular superlattice of the star-of-David CDW (Fig. 1(a)) can be seen regardless of the cleave-dependent surface state. Such topography is consistent with previous STM studies of bulk 1T-TaSe$_2$ surfaces [18,22].

Despite the fact that STM topographs of different cleaved domains of bulk 1T-TaSe$_2$ are nearly identical, their electronic structures vary dramatically from insulating to metallic behavior. The surface electronic structure remains constant within each macroscopic cleaving domain (typical size ≈ 100 μm) but differs strongly from one cleaving domain to another. The boundary between such domains is typically composed of one or more multi-layer step edges in our samples (a typical step height is ~3 nm; see Fig. S1 in the Supplementary Material (SM) [25] for more detail). Multi-layer steps between domains make stacking analysis at domain boundaries difficult and so here we will focus on the electronic structure obtained near the middle of different cleaving domains, where the surface is flat and homogeneous (e.g., Figs. 1(b)-1(d)).

We first discuss the insulating surface state of bulk 1T-TaSe$_2$. Fig. 1(b) shows a typical topograph of an insulating domain, but the electronic structure is best seen in Fig. 2. The black curve in Fig. 2(a) shows a d$I$/d$V$ point spectrum measured at the center of a star-of-David unit cell. d$I$/d$V$ curves obtained from STM spectroscopy represent local density of states (LDOS) and so Fig. 2(a) indicates the presence of a filled-state peak at $V_b$ = - 0.36 V, an energy gap around the Fermi level, and a pronounced empty-state peak at $V_b$ = 0.20 V. A second gap feature appears above the 0.20 V conduction band peak, beyond which higher-energy empty-state features rise up. Aside from small spatial variation in relative peak heights, this gapped electronic structure is spatially uniform within insulating cleavage planes (see Fig. S2 in the SM [25]). The spectral shape of the insulating bulk 1T-TaSe$_2$ surface state (including gap features and peak positions) shows striking similarity to STS measured on single-layer 1T-TaSe$_2$, and differs significantly from spectroscopy of bilayer and trilayer 1T-TaSe$_2$ (see Fig. S3 in the SM [25]) [5]. The overall features of the insulating surface state observed here are consistent with previously reported insulating surface states of bulk 1T-TaSe$_2$ [18,22].

To gain additional insight into the bulk insulating surface state, we performed STM differential conductance (d$I$/d$V$) mapping at a constant tip-sample separation (Figs. 2(b)-2(d)). At low energies (i.e., for low biases near the gap edges) the occupied-state d$I$/d$V$ map (Fig. 2(b)) and empty-state map (Fig. 2(c)) display nearly identical LDOS patterns with high-intensity LDOS appearing near the center of each star-of-David CDW unit cell. Lower-lying occupied states at $V_b$ < -0.3 V show a similar LDOS pattern, but higher-energy empty states at $V_b$ > 0.2 V (Fig. 2(d)) differ significantly and show a nearly inverted density pattern. To quantify the energy-dependent LDOS distribution of this phase we cross-correlate the d$I$/d$V$ maps with a reference map taken near the maximum of the occupied-state peak at $V_b$ = - 0.30 V (Fig. 2(b)). The similarity of the LDOS maps in the range -1 V < $V_b$ < 0.3 V can be seen by their strongly positive cross-correlation (blue) (a complete set of constant-height d$I$/d$V$ maps is shown in Fig. S4 in the SM [25]).

We next discuss weakly metallic surface states of bulk 1T-TaSe$_2$. As shown in Fig. 3(a), the d$I$/d$V$ spectrum of a weakly metallic surface hosts two strong LDOS peaks at $V_b$ = -0.27 V and $V_b$ = 0.22 V that bracket a low LDOS region around the Fermi level. Unlike the insulating state, a small zero-bias peak can be observed in the weakly metallic spectrum. This electronic structure is spatially homogeneous across entire cleavage domains (see Fig. S5 in the SM [25]). d$I$/d$V$ conductance mapping at constant tip-sample separation shows that the low-energy electronic LDOS maps (i.e., for -0.3 V ≤ $V_b$ ≤ 0.3 V) are dominated by the same CDW orbital which is concentrated near the center of each star-of-David cell (Figs. 3(b)-3(d)). Higher energy empty-state maps ($V_b$ > 0.5 V) exhibit very different LDOS patterns that are almost an inversion of the low-bias maps but that also show some additional structure (Fig. 3(e)) (a complete set of constant-height d$I$/d$V$ maps for the weakly metallic state can be seen in Fig. S6 in the SM [25]).

We lastly mention strongly metallic surface states observed in some cleavage domains of bulk 1T-TaSe$_2$. This surface state exhibits a much stronger (i.e., taller, wider) zero-bias peak than weakly metallic surface states (Fig. 4(a)), and is characterized by low-bias LDOS maps (-200 mV ≤ $V_b$ ≤ 200 mV) that are dominated by the same central CDW orbital as other low-bias 1T-TaSe$_2$ surface states (Fig. 4(b)-4(d)).

Figure 5 shows a direct comparison of STS performed on the three different classes of surface states observed for bulk 1T-TaSe$_2$. The behavior ranges from insulating to weakly metallic to strongly metallic, and the low-bias zoom-in spectra (Fig. 5(b)) clearly show that the metallic surface states always feature a zero-bias anomaly of varying strength. STM d$I$/d$V$ mapping allows us to clearly establish that the low-bias electronic structure of all three classes of surface states are dominated by the CDW orbital near the center of each star-of-David unit cell (Figs. 2, 3, and 4). This orbital is known to produce half-filled flat bands in single-layer 1T-TaSe$_2$ and single-layer 1T-TaS$_2$, and is thus expected to be responsible for the strong correlation physics seen in these materials [1-5].

## IV. Discussion

Our spectroscopic measurements of the different bulk 1T-TaSe$_2$ surface states suggest that they arise due to strong correlation physics. This can be seen foremost in the insulating case which is strikingly similar to the Mott insulator ground state of single-layer 1T-TaSe$_2$, and differs significantly from the bilayer and trilayer 1T-TaSe$_2$ multilayer systems (see Fig. S3 in the SM [25]). This implies that bulk 1T-TaSe$_2$ exhibits "decoupled-layer" surface electronic behavior reminiscent of the decoupled layer behavior seen in other layered materials such as graphite (e.g., the "graphene on graphite" behavior of refs. [26,27]) and which is attributed to stacking faults of the atomic lattice. Such decoupling might be further enhanced by stacking faults of the CDW lattice [1], resulting in nearly-isolated surface Mott insulating layers as observed here.

A natural explanation for the metallic surface states of 1T-TaSe$_2$ is that here the Mott surface layer is in better contact with the underlying bulk metallic layers [13,14] (most likely depending on the stacking-order-induced coupling [23]). In this case surface band renormalization due to stronger hybridization with the bulk layers explains the trend toward metallicity. Band renormalization of Mott/metal heterostructures has previously been predicted to yield a "three-peak" electronic structure with a sharp zero-bias Kondo peak as seen in the theoretically predicted spectral functions shown in Fig. 6 (adapted from ref. [28]). Here a heterostructure composed of stacked Mott insulating and metallic layers (Fig. 6 inset) was theoretically modelled by a periodic Anderson-Hubbard model [29] described by the following Hamiltonian [28]

$$H = t\sum_{\langle ij \rangle, \sigma} c_i^\dagger c_j + \sum_{i,z>0} U_{\text{Mott}}(n_{i\uparrow} - \frac{1}{2})(n_{i\downarrow} - \frac{1}{2}) + \sum_{i,z\leq 0} U_{\text{metal}}(n_{i\uparrow} - \frac{1}{2})(n_{i\downarrow} - \frac{1}{2}).$$

In this model the Mott (metal) layers have different $U/t$ ratios that are greater (less) than the critical threshold. When the coupling between the Mott and metal layers is the strongest (near the interface), a strong Kondo resonance appears between the Hubbard bands (blue curve in Fig. 6) due to metal-induced screening of local moments in the Mott layer. Decreasing the coupling between the Mott and metal layers quickly reduces the amplitude of the Kondo resonance peak (red curve) until the Kondo resonance is completely suppressed and the Mott insulating electronic structure is restored (black curve). Related behavior for single layers of Mott-insulating 1T-phase transition-metal dichalcogenide materials in contact with metallic 1H-phase materials has recently been observed experimentally [11,30,31] and theoretically [32]. The slave-rotor calculations in ref. [32] reveal fine structure in the Kondo lineshapes for these Mott/metal systems that resemble the zoom-in spectra shown in Fig. 5b.

Although the model shown in Fig. 6 qualitatively captures our observations, it requires *ad hoc U/t* values for the Mott/metal layers as well as adjustments in the effective coupling between Mott insulating and metal layers by varying distances from the interface. A more realistic model would better account for the microscopic details of our system such as the different coupling strengths between Mott insulating and metal layers due to different CDW stacking orders under the 1T-TaSe$_2$ surface (such stacking order has been shown theoretically [1,6-8] and experimentally [23,33-35] to significantly alter interlayer hopping strength). A periodic Anderson model or Hubbard model with this level of orbital complexity, however, has not yet been developed, but is probably necessary to fully explain the electronic structure of bulk 1T-TaSe$_2$/1T-TaS$_2$ and their surface states.

Bulk 1T-TaSe$_2$ is an ideal experimental platform for accessing a wide range of surfaces states exhibiting different coupling parameters. The main reason for this is that the star-of-David CDW transition temperature is very high for bulk 1T-TaSe$_2$ ($T_{CDW}$ = 473 K (above the room temperature) [13]). As a result, when the crystals are quenched to room temperature after growth, the three-dimensional star-of-David CDW structure rapidly freezes in and causes disordered CDW stacking along the out-of-plane direction (as shown in a recent Monte Carlo simulation [7]). Cleaving at room temperature thus naturally exposes a variety of different stacking sequences without allowing them to further relax. This is consistent with our observation of different surface electronic structure between cleaving domains, but uniform electronic structure within a single domain. In the case of the isostructural CDW material 1T-TaS$_2$, the star-of-David commensurate CDW transition sets in at a lower temperature of 180 K (below the room temperature) [13]. "Cold" cleaving of 1T-TaS$_2$ samples is thus required to freeze in CDW disorder, and such handling has indeed been found to yield previously unseen surface states [35,36]. In addition, since the bulk of 1T-TaSe$_2$ crystals very likely exhibits a comparable level of inhomogeneity to its surface states, we expect that further control of bulk CDW ordering, e.g., through delicately engineered thermal processes, might yield qualitatively different *bulk* electronic states (similar to the diverse surface states observed here) thus providing a means to realize new correlated phenomena even in the bulk form of this material [5,11,12,15-17].

## V. Conclusion

In conclusion, we have performed STM spectroscopic imaging of cleaved bulk 1T-TaSe$_2$ surfaces and have identified multiple domain-dependent correlated surface states, thus highlighting the central role of electron correlation in this material family. On insulating domains the bulk surface behavior strongly resembles the electronic structure of single-layer 1T-TaSe$_2$, suggesting the presence of a decoupled surface Mott layer. Metallic surface domains are observed to host zero-bias peaks of varying strength, pointing towards Kondo resonances arising from increased coupling between the surface Mott layer and underlying bulk metallic layers. The multiple 1T-TaSe$_2$ surface states observed here provide physical realizations of the periodic Anderson model [29,32,37,38] for different parameter regimes within the same crystal.

**Acknowledgements:**

This research was funded by the Director, Office of Science, Office of Basic Energy Sciences, Materials Sciences and Engineering Division, of the US Department of Energy

**References**

[1] P. Darancet, A. J. Millis, and C. A. Marianetti, Phys. Rev. B **90**, 045134 (2014) https://journals.aps.org/prb/abstract/10.1103/PhysRevB.90.045134.
[2] X.-L. Yu, D.-Y. Liu, Y.-M. Quan, J. Wu, H.-Q. Lin, K. Chang, and L.-J. Zou, Phys. Rev. B **96**, 125138 (2017) https://link.aps.org/doi/10.1103/PhysRevB.96.125138.
[3] S. Yi, Z. Zhang, and J.-H. Cho, Phys. Rev. B **97**, 041413(R) (2018) https://link.aps.org/doi/10.1103/PhysRevB.97.041413.
[4] Q. Zhang, L.-Y. Gan, Y. Cheng, and U. Schwingenschlögl, Phys. Rev. B **90**, 081103(R) (2014) https://link.aps.org/doi/10.1103/PhysRevB.90.081103.
[5] Y. Chen, W. Ruan, M. Wu, S. Tang, H. Ryu, H.-Z. Tsai, R. Lee, S. Kahn, F. Liou, C. Jia *et al.*, Nat. Phys. **16**, 218 (2020) https://doi.org/10.1038/s41567-019-0744-9.
[6] T. Ritschel, J. Trinckauf, K. Koepernik, B. Büchner, M. v. Zimmermann, H. Berger, Y. I. Joe, P. Abbamonte, and J. Geck, Nat. Phys. **11**, 328 (2015) https://www.nature.com/articles/nphys3267.
[7] S.-H. Lee, J. S. Goh, and D. Cho, Phys. Rev. Lett. **122**, 106404 (2019) https://link.aps.org/doi/10.1103/PhysRevLett.122.106404.
[8] T. Ritschel, H. Berger, and J. Geck, Phys. Rev. B **98**, 195134 (2018) https://link.aps.org/doi/10.1103/PhysRevB.98.195134.
[9] E. Tosatti and P. Fazekas, J. Phys. Colloq. **37**, 165 (1976) https://hal.archives-ouvertes.fr/jpa-00216540/document.
[10] P. Fazekas and E. Tosatti, Philos. Mag. B **39**, 229 (1979) https://www.tandfonline.com/doi/abs/10.1080/13642817908245359.
[11] W. Ruan, Y. Chen, S. Tang, J. Hwang, H.-Z. Tsai, R. L. Lee, M. Wu, H. Ryu, S. Kahn, F. Liou *et al.*, Nat. Phys. **17**, 1154 (2021) https://doi.org/10.1038/s41567-021-01321-0.
[12] Y. Chen, W.-Y. He, W. Ruan, J. Hwang, S. Tang, R. L. Lee, M. Wu, T. Zhu, C. Zhang, and H. Ryu, arXiv preprint arXiv:2202.07224 (2022) https://arxiv.org/abs/2202.07224.
[13] J. A. Wilson, F. J. Di Salvo, and S. Mahajan, Adv. Phys. **24**, 117 (1975) https://www.tandfonline.com/doi/abs/10.1080/00018737500101391.
[14] F. J. Di Salvo, R. G. Maines, J. V. Waszczak, and R. E. Schwall, Solid State Commun. **14**, 497 (1974) http://www.sciencedirect.com/science/article/pii/0038109874909752.
[15] K. T. Law and P. A. Lee, Proc. Natl. Acad. Sci. U.S.A. **114**, 6996 (2017) https://www.ncbi.nlm.nih.gov/pubmed/28634296.
[16] W.-Y. He, X. Y. Xu, G. Chen, K. T. Law, and P. A. Lee, Phys. Rev. Lett. **121**, 046401 (2018) https://link.aps.org/doi/10.1103/PhysRevLett.121.046401.
[17] X. Y. Xu, K. T. Law, and P. A. Lee, Phys. Rev. Lett. **122**, 167001 (2019) https://link.aps.org/doi/10.1103/PhysRevLett.122.167001.
[18] O. Shiino, T. Endo, W. Yamaguchi, H. Sugawara, K. Kitazawa, and T. Hasegawa, Applied Physics A **66**, S175 (1998) https://doi.org/10.1007/s003390051125.
[19] J.-J. Kim, W. Yamaguchi, T. Hasegawa, and K. Kitazawa, Phys. Rev. B **50**, 4958 (1994) https://journals.aps.org/prb/abstract/10.1103/PhysRevB.50.4958.
[20] B. Giambattista, C. G. Slough, W. W. McNairy, and R. V. Coleman, Phys. Rev. B **41**, 10082 (1990) https://link.aps.org/pdf/10.1103/PhysRevB.41.10082.
[21] L. Perfetti, A. Georges, S. Florens, S. Biermann, S. Mitrovic, H. Berger, Y. Tomm, H. Hochst, and M. Grioni, Phys. Rev. Lett. **90**, 166401 (2003) https://journals.aps.org/prl/abstract/10.1103/PhysRevLett.90.166401.
[22] S. Colonna, F. Ronci, A. Cricenti, L. Perfetti, H. Berger, and M. Grioni, Phys. Rev. Lett. **94**, 036405 (2005) https://journals.aps.org/prl/abstract/10.1103/PhysRevLett.94.036405.

[23] W. Zhang, Z. Wu, K. Bu, Y. Fei, Y. Zheng, J. Gao, X. Luo, Z. Liu, Y.-P. Sun, and Y. Yin, Phys. Rev. B **105**, 035110 (2022) https://link.aps.org/doi/10.1103/PhysRevB.105.035110.
[24] R. Brouwer and F. Jellinek, Physica B+C **99**, 51 (1980) http://www.sciencedirect.com/science/article/pii/0378436380902090.
[25] See Supplemental Material at [URL will be inserted by publisher] for more STM/STS data.
[26] G. Li, A. Luican, and E. Y. Andrei, Phys. Rev. Lett. **102**, 176804 (2009) https://link.aps.org/doi/10.1103/PhysRevLett.102.176804.
[27] G. Li and E. Y. Andrei, Nat. Phys. **3**, 623 (2007) https://doi.org/10.1038/nphys653.
[28] R. W. Helmes, T. A. Costi, and A. Rosch, Phys. Rev. Lett. **101**, 066802 (2008) https://link.aps.org/doi/10.1103/PhysRevLett.101.066802.
[29] T. Schork and S. Blawid, Phys. Rev. B **56**, 6559 (1997) https://link.aps.org/doi/10.1103/PhysRevB.56.6559.
[30] V. Vaňo, M. Amini, S. C. Ganguli, G. Chen, J. L. Lado, S. Kezilebieke, and P. Liljeroth, Nature **599**, 582 (2021) https://doi.org/10.1038/s41586-021-04021-0.
[31] M. Liu, J. Leveillee, S. Lu, J. Yu, H. Kim, C. Tian, Y. Shi, K. Lai, C. Zhang, F. Giustino *et al.*, Sci. Adv. **7**, eabi6339 (2021) https://doi.org/10.1126/sciadv.abi6339.
[32] C. Chen, I. Sodemann, and P. A. Lee, Phys. Rev. B **103**, 085128 (2021) https://link.aps.org/doi/10.1103/PhysRevB.103.085128.
[33] L. Ma, C. Ye, Y. Yu, X. F. Lu, X. Niu, S. Kim, D. Feng, D. Tomanek, Y. W. Son, X. H. Chen *et al.*, Nat. Commun. **7**, 10956 (2016) https://www.nature.com/articles/ncomms10956.
[34] D. Cho, S. Cheon, K. S. Kim, S. H. Lee, Y. H. Cho, S. W. Cheong, and H. W. Yeom, Nat. Commun. **7**, 10453 (2016) https://www.nature.com/articles/ncomms10453.
[35] C. J. Butler, M. Yoshida, T. Hanaguri, and Y. Iwasa, Nat. Commun. **11**, 2477 (2020) https://doi.org/10.1038/s41467-020-16132-9.
[36] C. J. Butler, M. Yoshida, T. Hanaguri, and Y. Iwasa, Phys. Rev. X **11**, 011059 (2021) https://link.aps.org/doi/10.1103/PhysRevX.11.011059.
[37] C. Noce, Physics Reports **431**, 173 (2006) https://www.sciencedirect.com/science/article/pii/S0370157306001621.
[38] P. W. Anderson, Physical Review **124**, 41 (1961) https://link.aps.org/doi/10.1103/PhysRev.124.41.

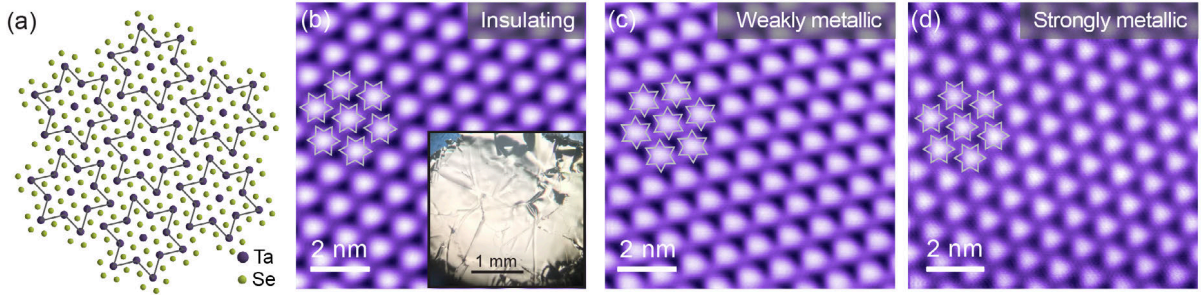

Fig. 1. Charge density wave at bulk 1T-TaSe$_2$ surface. (a) Schematic of star-of-David CDW supercells in 1T-TaSe$_2$ (top view). (b)-(d) STM topographs show star-of-David CDW patterns for different bulk 1T-TaSe$_2$ surface states exhibiting (b) insulating, (c) weakly metallic, and (d) strongly metallic electronic ground states. Star-of-David CDW unit cells are outlined (white line). ((b) and (c) $V_b$ = -1 V, $I_t$ = 10 pA; (d) $V_b$ = -1 V, $I_t$ = 5 pA). Inset in (b) shows an optical image of a typical cleaved surface of bulk 1T-TaSe$_2$.

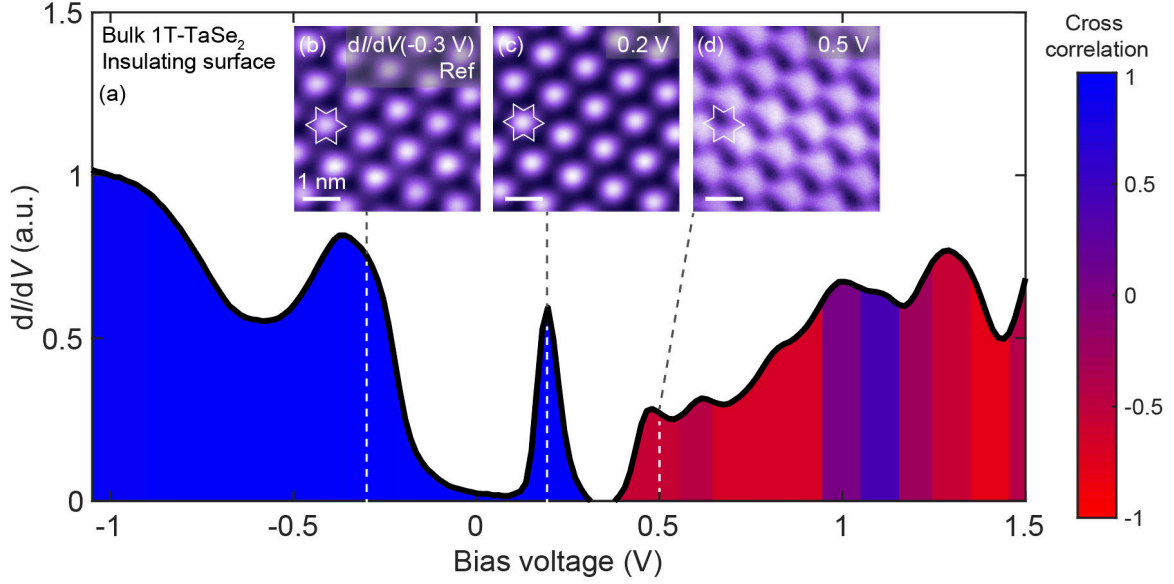

Fig. 2. Electronic structure of the insulating surface state for bulk 1T-TaSe$_2$. (a) STM d$I$/d$V$ spectrum shows insulating surface state behavior for bulk 1T-TaSe$_2$ ($f$ = 401 Hz, $I_t$ = 20 pA, $V_{RMS}$ = 20 mV). (b)-(d) d$I$/d$V$ maps of the insulating surface state for (b) $V_b$ = -0.3 V, (c) $V_b$ = 0.2 V, and (d) $V_b$ = 0.5 V ($f$ = 401 Hz, $V_{RMS}$ = 30 mV). The low-bias electronic states [(b) and (c)] are dominated by a CDW orbital concentrated near the center of each star-of-David CDW supercell. The same star-of-David CDW unit cell is outlined in each map (white line). Color shows cross-correlation of d$I$/d$V$ maps at different energies with the reference map shown in (b).

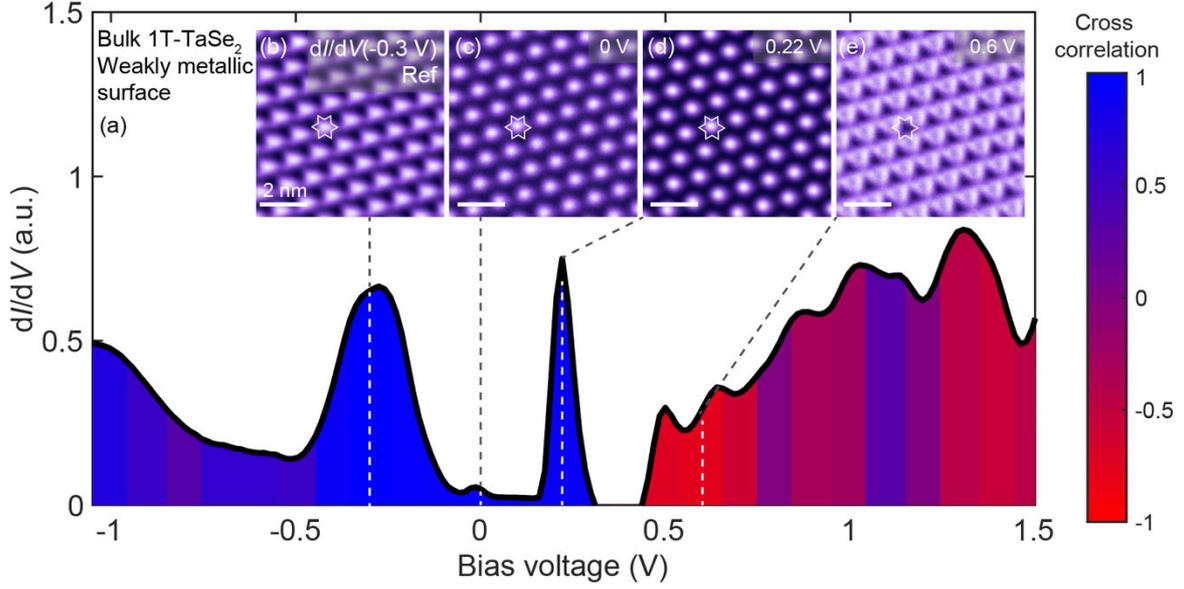

Fig. 3. Electronic structure of the weakly metallic surface state for bulk 1T-TaSe$_2$. (a) STM d$I$/d$V$ spectrum shows weakly metallic surface state behavior with a small zero-bias peak for bulk 1T-TaSe$_2$ ($f$ = 401 Hz, $I_t$ = 20 pA, $V_{RMS}$ = 20 mV). (b)-(e) Constant-height d$I$/d$V$ maps of the weakly metallic surface state for (b) $V_b$ = -0.3 V, (c) $V_b$ = 0 V, (d) $V_b$ = 0.22 V, and (e) $V_b$ = 0.6 V ($f$ = 401 Hz, $V_{RMS}$ = 20 mV). The low-bias electronic states [(b)-(d)] are all dominated by the same CDW orbital concentrated near the center of each star-of-David CDW unit cell. Color shows cross-correlation of d$I$/d$V$ maps at different energies with the reference map shown in (b).

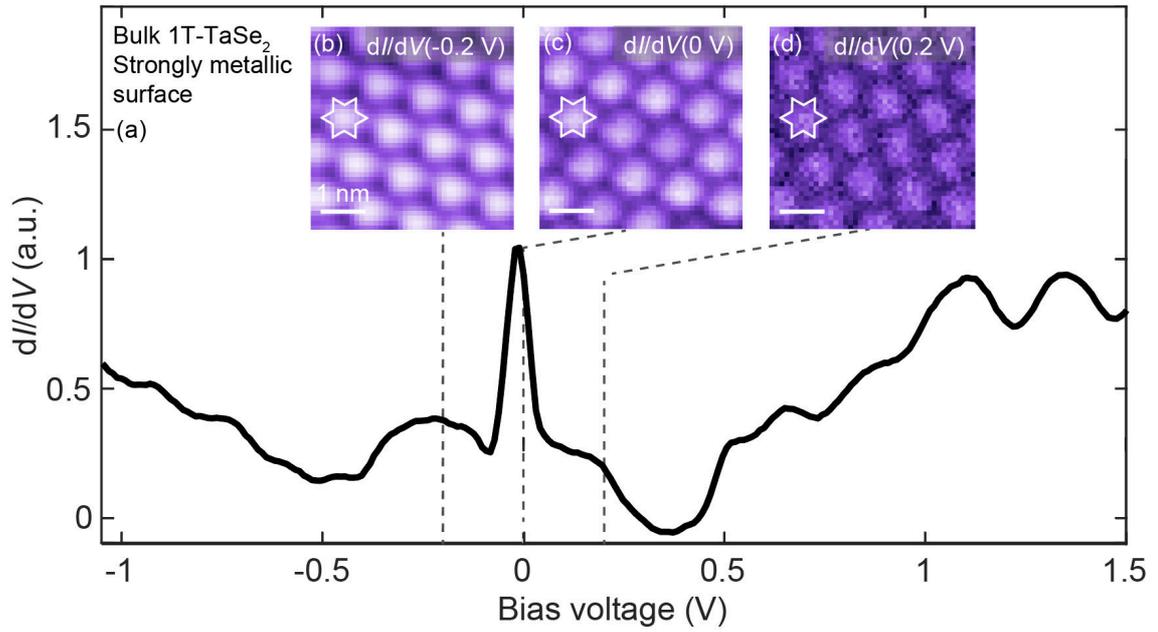

Fig. 4. Electronic structure of the strongly metallic surface state for bulk 1T-TaSe$_2$. (a) STM d$I$/d$V$ spectrum shows strongly metallic surface state with a large zero-bias peak for bulk 1T-TaSe$_2$ ($f$ = 401 Hz, $I_t$ = 20 pA, $V_{RMS}$ = 20 mV). (b)-(d) Constant-current d$I$/d$V$ maps of the strongly metallic surface state for (b) $V_b$ = -0.2 V, (c) $V_b$ = 0 V, and (d) $V_b$ = 0.2 V ($f$ = 401 Hz, $V_{RMS}$ = 2 mV). The low-bias electronic states [(b)-(d)] are all dominated by the same CDW orbital concentrated near the center of each star-of-David CDW unit cell.

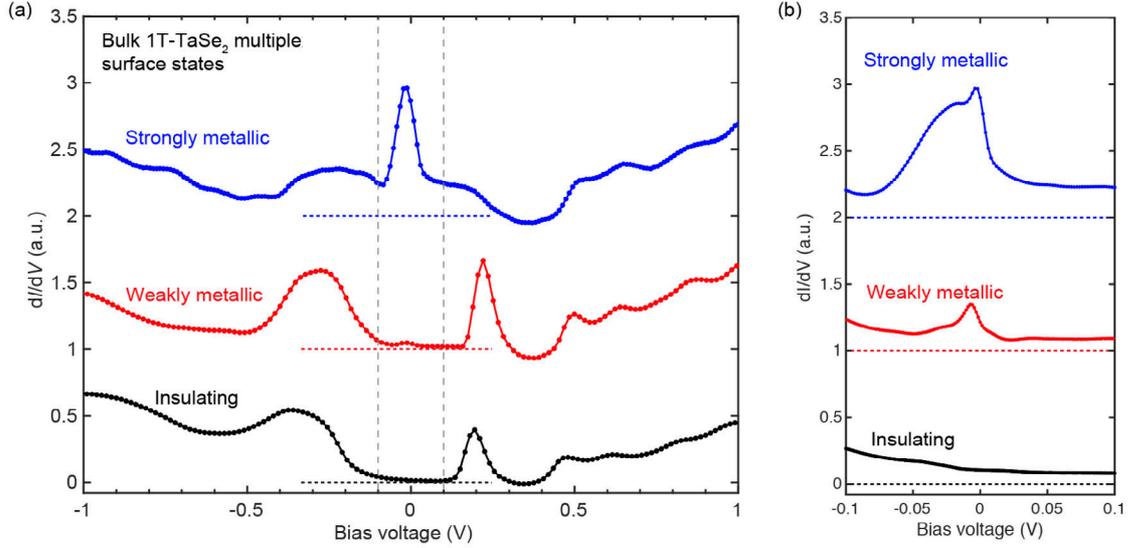

Fig. 5. Different surface states for bulk 1T-TaSe$_2$. (a) Wide-bias STM d$I$/d$V$ spectra show different surface electronic structure for bulk 1T-TaSe$_2$. The surface behavior ranges from fully insulating to weakly metallic (i.e., having a small zero-bias peak) and to strongly metallic (i.e., having a large zero-bias peak) ($f$ = 401 Hz, $I_t$ = 20 pA, $V_{RMS}$ = 20 mV). Curves are shifted vertically for viewing (horizontal dashed lines mark d$I$/d$V$ = 0). (b) Low-bias zoom-in spectra of the different possible surface behaviors for bulk 1T-TaSe$_2$ show zero-bias peaks for metallic surface states and a featureless LDOS near $V_b$ = 0 for the insulating surface state ($f$ = 401 Hz, $I_t$ = 20 pA, $V_{RMS}$ = 2 mV for metallic surface states, $V_{RMS}$ = 5 mV for insulating surface state). Curves are shifted vertically for viewing (horizontal dashed lines mark d$I$/d$V$ = 0).

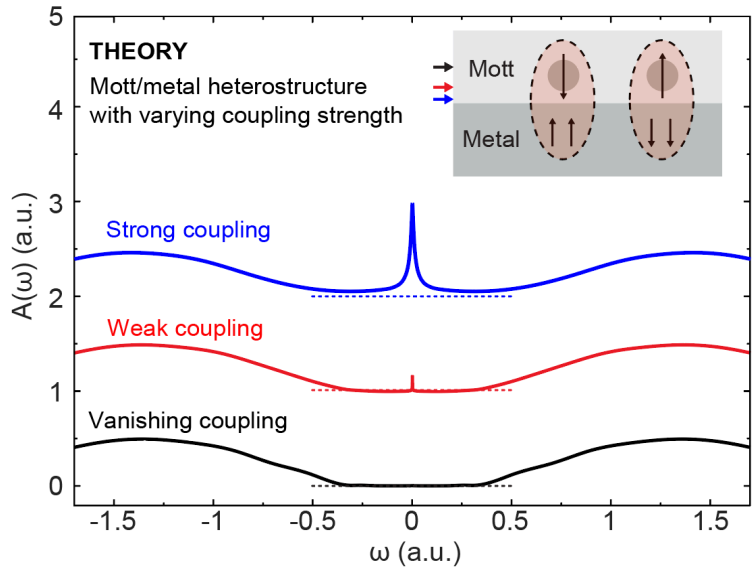

Fig. 6. Theoretical spectral functions of a Mott/metal heterostructure adapted from Ref. [28]. Theoretical spectral functions for strong coupling (blue), weak coupling (red), and no coupling (black) between the Mott and metal layers. The strength of the zero-bias Kondo peak decreases quickly with reduced coupling. Inset shows a representative sketch of the Mott/metal heterostructure where spins in the Mott layers near the interface are screened by itinerant electrons in the metal layers. Colored arrows show the locations where the spectral functions were calculated.

Supplementary Materials:

# Observation of a multitude of correlated states at the surface of bulk 1T-TaSe$_2$ crystals

Yi Chen[†], Wei Ruan[†], Jeffrey D. Cain, Ryan L. Lee, Salman Kahn, Caihong Jia, Alex Zettl, Michael F. Crommie[*]

[†] These authors contributed equally to this work.

[*]e-mail: crommie@berkeley.edu

**Table of Contents**

**Experimental set up**

**1. Sample growth**

Bulk 1T-TaSe$_2$ was grown by chemical vapor transport using iodine as a transport agent. Stoichiometric amounts of Ta and Se were sealed under high vacuum in a quartz ampoule and heated to 950 °C for a period of 1-2 weeks. After the growth period the 1T-TaSe$_2$ crystals were rapidly quenched to room temperature in ice water to maintain the 1T structure. The crystal structure of 1T-TaSe$_2$ was verified by X-ray diffraction measurements of bulk samples as well as transport measurements of exfoliated thin-film devices, both showing pure 1T signals.

**2. STM/STS measurements**

STM and STS measurements were performed in a low-temperature ultrahigh-vacuum STM system (CreaTec) at $T = 5$ K. Electrochemically etched tungsten tips were calibrated on a Au(111) surface before measurements. The bulk 1T-TaSe$_2$ crystals were cleaved under UHV conditions at room temperature before all STM measurements. Thermal annealing was minimized to avoid possible structural transitions. d$I$/d$V$ spectra were collected using standard lock-in techniques ($f = 401$ Hz). d$I$/d$V$ mapping was performed in constant-height mode (i.e., with the feedback loop open) except for Fig. 4 (where the constant-current mode is used).

**Supplementary Figures**

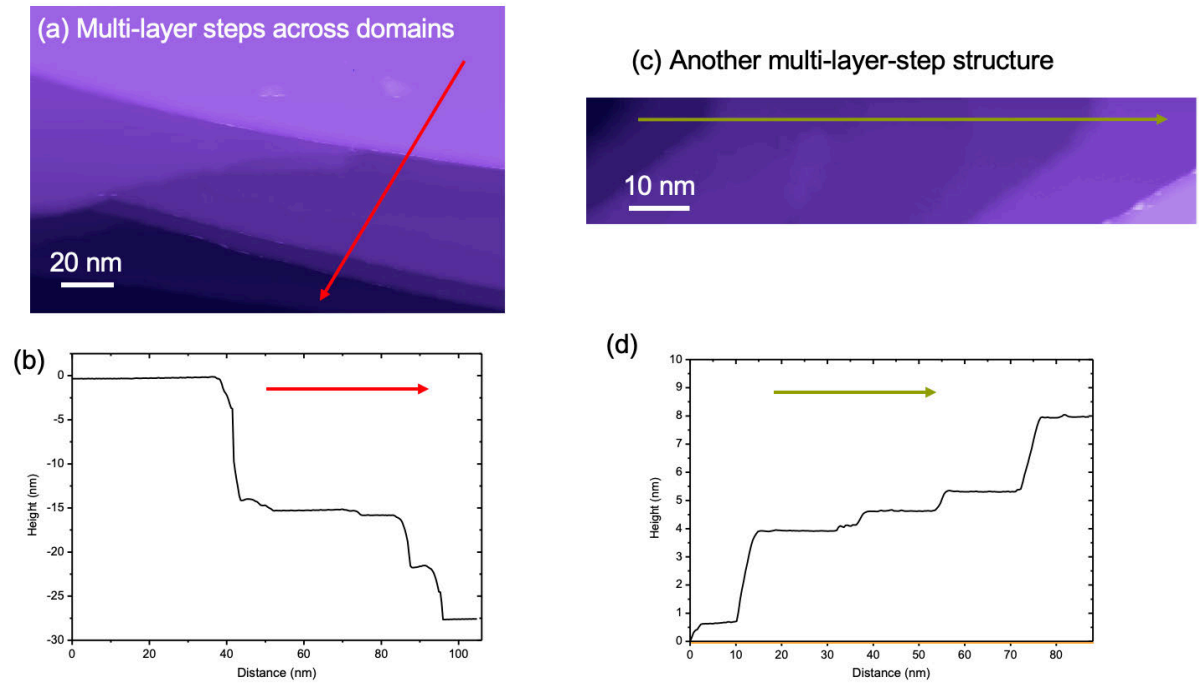

**Fig. S1. Multi-layer step structures across different domains at bulk 1T-TaSe$_2$ surface.** (a) and (b) STM topograph of a multi-step structure between different domains at bulk 1T-TaSe$_2$ surface after cleaving. The red arrow shows where the linecut in (b) was taken ($V_b$ = -1 V, $I_t$ = 5 pA). (c) and (d) Similar to (a) and (b) but in a different region ($V_b$ = -1 V, $I_t$ = 5 pA).

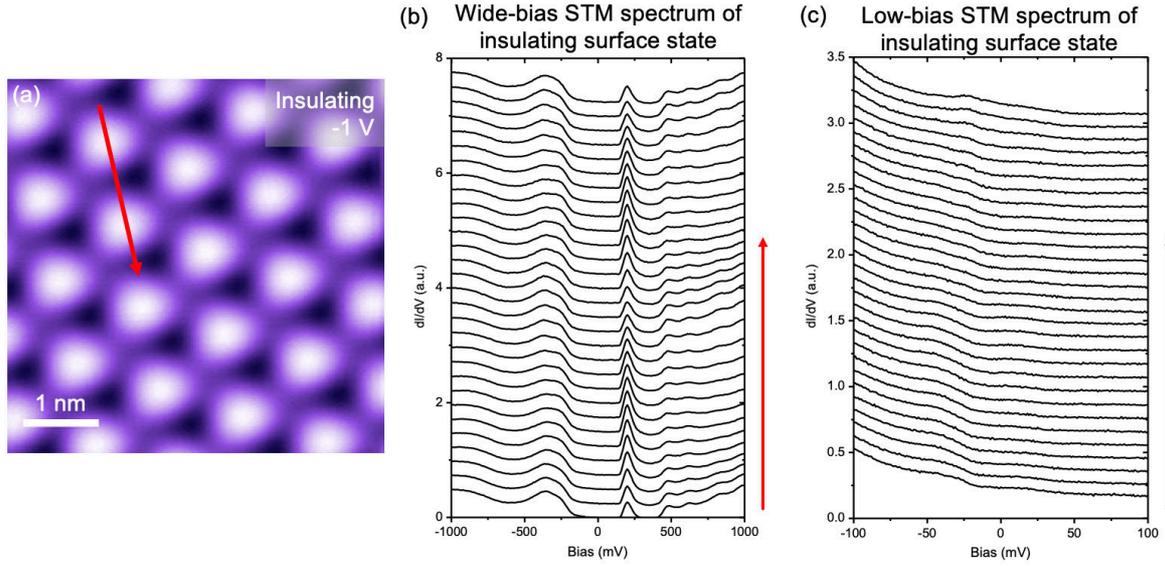

**Fig. S2. Spatially homogeneous surface state in an insulating domain of bulk 1T-TaSe$_2$.** (a) STM topograph of an insulating domain of bulk 1T-TaSe$_2$. The red arrow shows where STS line spectra were acquired ($V_b$ = -1 V, $I_t$ = 5 pA). (b) Wide-bias STS spectra acquired along the red arrow show spatially-uniform gapped electronic structure ($f$ = 401 Hz, $I_t$ = 50 pA, $V_{RMS}$ = 20 mV). (c) Low-bias STS spectra acquired along the same arrow show spatially-uniform featureless electronic structure near $V_b$ = 0 ($f$ = 401 Hz, $I_t$ = 20 pA, $V_{RMS}$ = 5 mV).

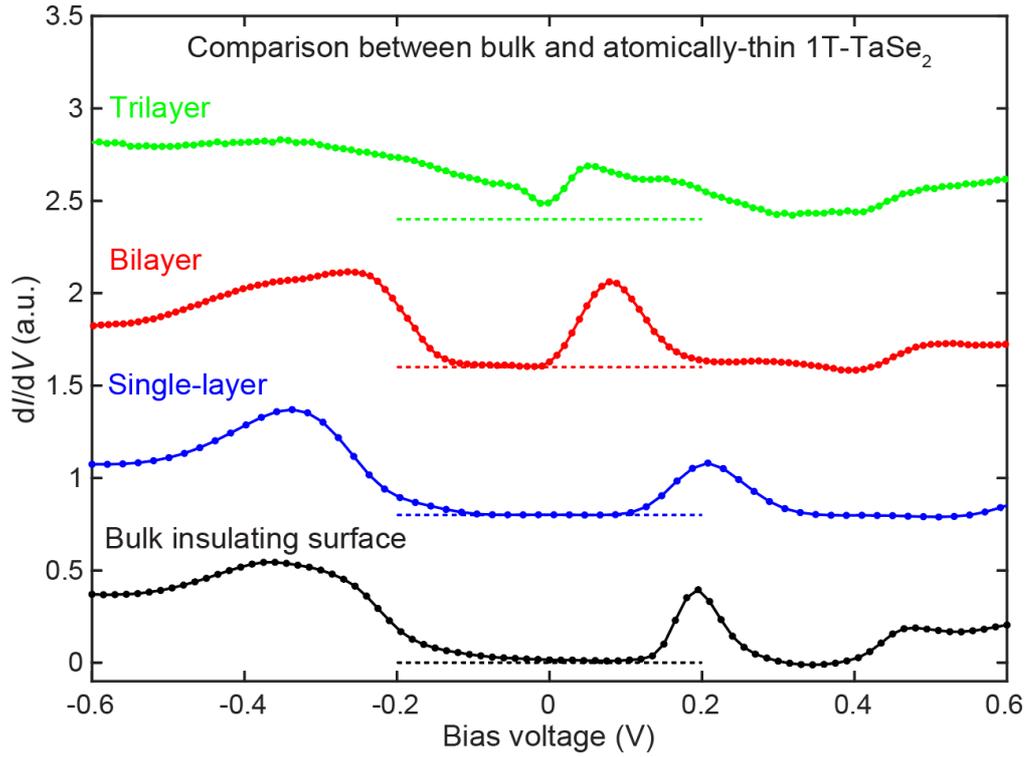

**Fig. S3. STM spectra taken on the insulating surface state of bulk 1T-TaSe$_2$ compared with single-layer, bilayer, and trilayer 1T-TaSe$_2$.**

The spectral features of the insulating surface state of bulk 1T-TaSe$_2$ strongly resemble single-layer 1T-TaSe$_2$. The STM spectra of bilayer and trilayer 1T-TaSe$_2$ have very different gap sizes and spectral shapes ($f$ = 401 Hz, $V_{RMS}$ = 20~30 mV).

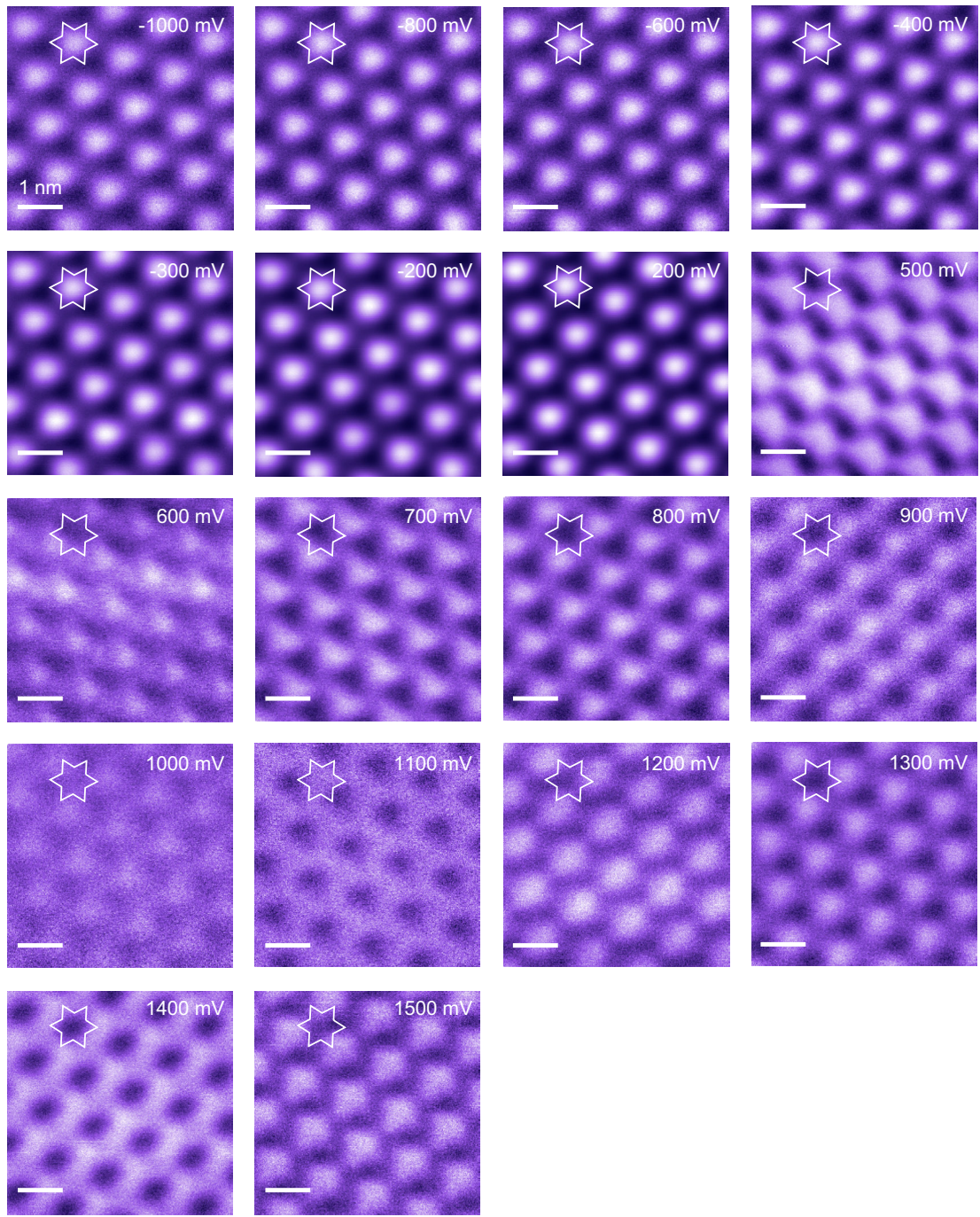

**Fig. S4.** Constant-height d$I$/d$V$ conductance maps of the insulating surface state of bulk 1T-TaSe$_2$ at different biases ($f$ = 401 Hz, $V_{RMS}$ = 30 mV).

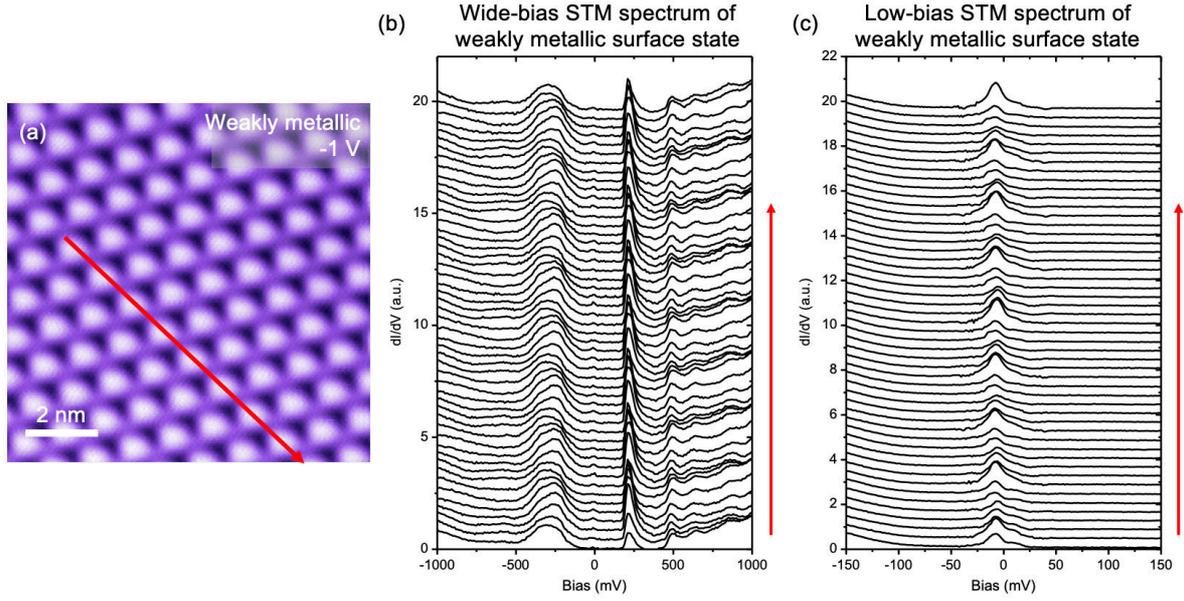

**Fig. S5. Spatially homogeneous surface state in a weakly metallic domain of bulk 1T-TaSe$_2$.**

(a) STM topograph of a weakly metallic domain of bulk 1T-TaSe$_2$. The red arrow shows where STS line spectra were acquired ($V_b$ = -1 V, $I_t$ = 5 pA). (b) Wide-bias STS spectra acquired along the red arrow show spatially-uniform electronic structure ($f$ = 401 Hz, $I_t$ = 30 pA, $V_{RMS}$ = 10 mV). (c) Low-bias STS spectra acquired along the same arrow show mostly spatially uniform electronic structure except for a small periodic modulation in the amplitude of the zero-bias peak (period ~ 1.2 nm) ($f$ = 401 Hz, $I_t$ = 50 pA, $V_{RMS}$ = 1 mV).

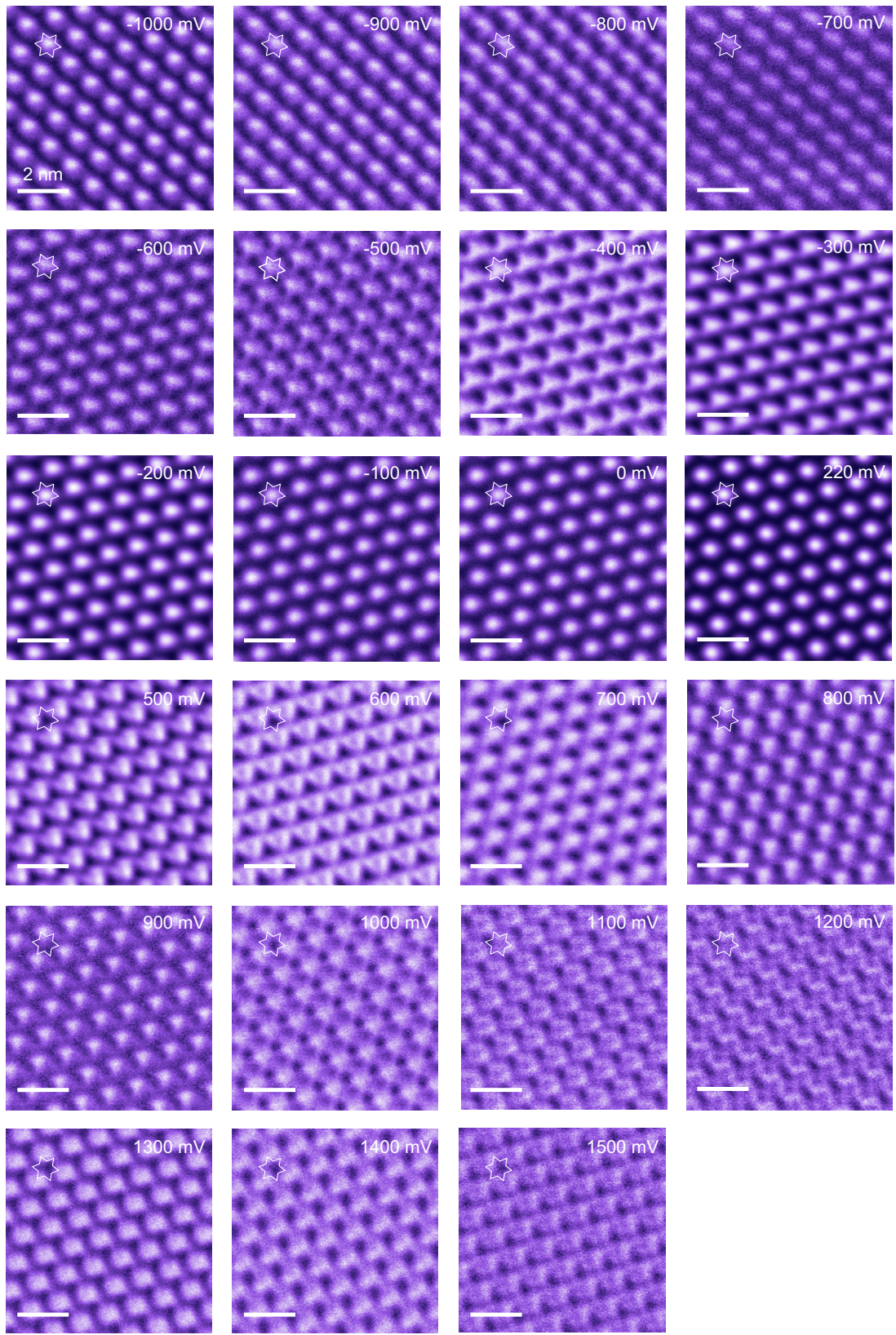

Fig. S6. Constant-height d$I$/d$V$ conductance maps of the weakly metallic surface state in bulk 1T-TaSe$_2$ at different biases ($f$ = 401 Hz, $V_{RMS}$ = 20 mV).

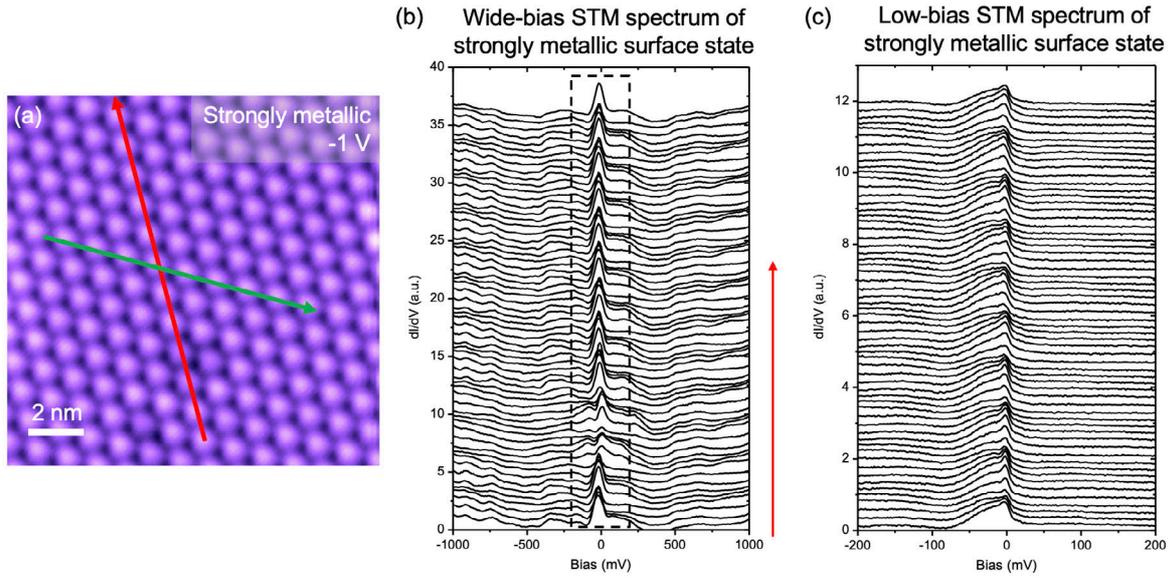

**Fig. S7. Homogeneous surface state in a strongly metallic domain of bulk 1T-TaSe$_2$.**
(a) STM topograph of a strongly metallic domain of bulk 1T-TaSe$_2$. The red (green) arrow shows where the wide-bias (low-bias) STS spectra were acquired ($V_b$ = -1 V, $I_t$ = 5 pA). (b) Wide-bias STS spectra acquired along the red arrow show spatially uniform electronic structure ($f$ = 401 Hz, $I_t$ = 30 pA, $V_{RMS}$ = 20 mV). The dashed box indicates the bias range used in (c). (c) Low-bias STS spectra acquired along the green arrow show mostly spatially uniform electronic structure except for small periodic modulation in the peak amplitude ($f$ = 401 Hz, $I_t$ = 30 pA, $V_{RMS}$ = 1 mV).